\documentclass[showpacs,amssymb,preprintnumbers,
nofootinbib, superscriptaddress ]{revtex4}

\usepackage{amsmath}
\usepackage{graphicx}
\usepackage{latexsym}
\usepackage{amsfonts}
\usepackage{url,hyperref}
\usepackage{bm}

\begin{document}

\begin{figure}[t]
\vspace{-1.4cm}
\hspace{-16.25cm}
\end{figure}

\newcommand{\vp}{\varphi}
\newcommand{\nn}{\nonumber\\}
\newcommand{\beq}{\begin{equation}}
\newcommand{\eeq}{\end{equation}}
\newcommand{\bed}{\begin{displaymath}}
\newcommand{\eed}{\end{displaymath}}
\def\bea{\begin{eqnarray}}
\def\eea{\end{eqnarray}}
\newcommand{\veps}{\varepsilon}

\title{Fermion excitations of a tense brane black hole}
\author{H.~T.~Cho}
\email{htcho\_at\_mail.tku.edu.tw}
\affiliation{Department of Physics, Tamkang University, Tamsui, Taipei, Taiwan, Republic of China}
\author{A.~S.~Cornell}
\email[Email: ]{cornell\_at\_ipnl.in2p3.fr}
\affiliation{Institut de Physique Nucl{\'e}aire, Universit{\'e} Claude Bernard (Lyon 1), 4 rue Enrico Fermi, F - 69622 - Villeurbanne, France}
\author{Jason~Doukas}
\email[Email: ]{j.doukas\_at\_physics.unimelb.edu.au}
\affiliation{School of Physics, University of Melbourne, Parkville, Victoria 3010, Australia.}
\author{Wade~Naylor}
\email[Email: ]{naylor\_at\_se.ritsumei.ac.jp}
\affiliation{Department of Physics, Ritsumeikan University, Kusatsu, Shiga 525-8577, Japan}

\begin{abstract}
By finding the spinor eigenvalues for a single deficit angle $(d-2)$-sphere, we derive
the radial potential for fermions on a $d$-dimensional black hole  background that is embedded on a codimension two brane with conical singularity, where the deficit angle is related to the brane tension. From this we obtain the quasi-normal mode spectrum for bulk fermions on such a background. As a byproduct of our method, this also gives a rigorous proof for integer spin fields on the deficit 2-sphere.
\end{abstract}

\date{October 29, 2007}
\maketitle

\section{Introduction}
\par Much has been said about black holes (BHs) in large extra-dimensional scenarios recently, e.g., see \cite{GiddRev} and the references therein. Such theories lead to the intriguing possibility that the LHC could actually even produce BHs, with various particle species being emitted by Hawking radiation or via classical BH excitations known as quasi-normal modes (QNMs). Although it is not entirely clear as to whether or not detection of such processes is feasible, work still remains to be done on various issues.
For example, the effect of the brane tension for BHs with large extra dimensions has largely been ignored, because of the obvious difficulty of how to embed a BH onto a brane.

\par However, recently the work of Kaloper \& Kiley \cite{KK}, inspired from codimension-two braneworld models, presented the following metric for a black hole residing on a tensional 3-brane embedded in a six-dimensional spacetime:
\beq
\label{metric}
 \mathrm{d}s^{2}=-f(r)dt^{2}
+\frac{dr^{2}}{f(r)}+ r^2 d\Omega_4^2 \ \ , \ \ \ \
f(r)=1-\left(\frac{r_{H}}{r}\right)^{3}
 \eeq
 where the radius of the horizon is given by
\beq\label{eqn:tenseradius}
r_H=\left(\frac{\mu}{b}\right)^{1/3} \,\qquad\qquad\, \mu \equiv \frac{M_{BH}}{4\pi^2M_*^4}
\eeq
and $M_{BH}$ is the mass of the black hole. The parameter $b$ is a measure of the conical
deviation from a perfect sphere and has the following angle element:
\beq
 d\Omega_4^2 = d\theta^{2}_3+\sin^{2}\theta_3
\left(d\theta_2^2+\sin^{2}\theta_2 \left(d\theta_1 ^{2}+
b^{2}\sin^{2}\theta_1 d\phi ^{2}\right) \right), \ \ \ \ 0 < b \le 1 \,
\label{anglel}
\eeq
where for $b=1$ this is the line element of the unit sphere $S^4$ and corresponds to zero brane tension. It may be worth mentioning that the location of the deficit angle is arbitrary and that it is possible to consider more than one deficit angle, where such cases may be of interest to the fermion generation puzzle re-expressed as a higher-dimensional problem, see \cite{Singleton} and the references therein.

For a non-vanishing brane tension the parameter $b<1$ is a measure of the deficit angle about an axis parallel to the 3-brane in the angular direction $\phi$, such that the canonically normalized angle $\phi'=\phi/b$ runs over the interval $[0,2\pi/b]$. It can be expressed in term of the brane tension $\lambda$ as:
\beq
\label{eqb} b = 1 - \frac{\lambda}{4\pi M_*^4} \,,
\eeq
where $M_*$ is the fundamental Planck constant of six-dimensional gravity. As can be seen the tension of the brane modifies the radius of the horizon, namely it increases with increasing tension ($b\to 0$).

The field equations for the scalar perturbations are discussed in
\cite{Dai,CWS,KK} and similarly for graviton and electromagnetic
perturbations in \cite{Siopsis}. The emission rates were
calculated in \cite{Dai} for scalar, gauge boson and graviton
fields, where numerical methods were used to solve the angular
eigenvalue equations. Also, the scalar QNMs are found in
\cite{CWS} by a perturbative expansion in powers of $(1-b)$ and
hence are only valid for branes with small tensions. Recently the
QNMs for scalar and gravitational perturbations have been found
exactly, based on eigenflow arguments for integral $1/b$
\cite{Siopsis}.

In this article we shall fill the gap by presenting results for the spin-half QNMs. Furthermore, the method we shall apply to spinors (by choosing azimuthal eigenvalue $m=\pm 1/2,\pm 3/2, \dots $) also applies directly to the other perturbations  (choosing eigenvalue $m=0, \pm 1,\pm /2, \dots $) and leads to a rigorous proof of the 2-sphere angular eigenvalue with no assumptions made on the form of $b$.

\section{Spinor Fields}

\par The above works have only dealt with integer spin fields and in terms of phenomenology spin half fields are also important. For generality we shall begin our analysis by supposing a background metric which is $d$-dimensional and spherically symmetric, as given by:
\begin{equation}
ds^{2}=-f(r)dt^{2}+h(r)dr^{2}+r^{2}d\Omega^{2}_{d-2} ,
\end{equation}
where now $d \Omega^{2}_{d-2}$ denotes the metric for the $(d-2)$-dimensional deficit sphere which in six dimensions has line element given by equation (\ref{anglel}).

\par Then under a conformal transformation:
\begin{eqnarray}
g_{\mu\nu} & \rightarrow & \overline{g}_{\mu\nu}=\Omega^{2}g_{\mu\nu} , \\
\psi & \rightarrow & \overline{\psi}=\Omega^{-(d-1)/2}\psi , \\
\gamma^{\mu}\nabla_{\mu}\psi & \rightarrow & \Omega^{(d+1)/2} \overline{\gamma}^{\mu}\overline{\nabla}_{\mu}\overline{\psi} ,
\end{eqnarray}
where we shall take $\Omega=1/r$, the metric becomes:
\beq
d\overline{s}^{2} = -\frac{f}{r^{2}}dt^{2} + \frac{h}{r^{2}}dr^{2} + d\Omega^{2}_{d-2} , \qquad\qquad \mathrm{where} \qquad \overline{\psi}=r^{(d-1)/2}\psi . \\
\eeq
Since the $t-r$ part and the $(d-2)$-sphere part of the metric are completely separated, one can write the Dirac equation in the form:
\begin{eqnarray}
& \overline{\gamma}^{\mu}\overline{\nabla}_{\mu}\overline{\psi}=0 , & \nonumber\\
& \Rightarrow \left[ \left( \overline{\gamma}^{t}\overline{\nabla}_{t} + \overline{\gamma}^{r}\overline{\nabla}_{r} \right) \otimes 1 \right] \overline{\psi} + \left[ \overline{\gamma}^{5} \otimes \left( \overline{\gamma}^{a} \overline{\nabla}_{a}\right)_{S_{d-2}} \right] \overline{\psi} = 0 , &
\end{eqnarray}
where $(\overline{\gamma}^{5})^{2}=1$. Note that from this point on we shall change our notation by omitting the bars.

\par The problem now is to find the eigenvalue for the projected $\chi_{l}^{(\pm)}$, which are the  eigenspinors for the deficit $(d-2)$-sphere, that is:
\begin{equation}
\left( \gamma^{a}\nabla_{a} \right)_{S_{d-2}}\chi_{l}^{(\pm)} = \pm i \kappa \chi_{l}^{(\pm)} \,.
\end{equation}
The separation follows exactly as in \cite{CCDN} and leads to the following radial Schr\"odinger-like equation in the tortoise coordinate $r_{*}$:
\beq
\left(-\frac{d^2}{dr_{*}^{2}}+V_{1}\right)G=E^{2}G \,\,\, ,
\label{gequation}
\eeq
where $G$ is the upper component of the two-component spinor \cite{CCDN} and the potential is given by:
\begin{eqnarray}
V_1(r) &=&\kappa^2 {f\over r^2}+\kappa
f\frac{d}{dr}\left[\frac{\sqrt{f}}{r}\right] \,,
\label{V1}
\end{eqnarray}
where we have set $h(r)=1/f(r)$ for the Schwarzschild case, and
the eigenvalue, $\kappa$, shall be determined in the next section.
\section{Deficit $2$-sphere}

\par We shall first consider the case of a deficit $2$-sphere, whereby we can generate results for the $(d-2)$-sphere, using eigenflow arguments much like in \cite{Siopsis}. First of all let's suppose the metric of the deficit two sphere be
\begin{equation}
ds^{2}=d\theta^{2}+b^{2}\sin^{2}\theta\ d\phi^{2}\,,
\end{equation}
where $b$ is a positive real number and $b=1$ represents a regular two sphere.
The Dirac operator is then given by
\begin{equation}
\gamma^{a}\nabla_{a}\psi
=\gamma^{a}{e_{a}}^{\mu}\left(\partial_{\mu}+\Gamma_{\mu}\right)\psi,
\end{equation}
where the spin connection $\Gamma_{\mu}$ is given in terms of the
zweibein ${e_{a}}^{\mu}$ and its inverse,
\begin{equation}
\Gamma_{\mu}=\frac{1}{8}\left[\gamma^{a},\gamma^{b}\right]
{e_{a}}^{\nu}\left(\partial_{\mu}e_{b\nu}-\Gamma^{\alpha}_{\mu\nu}
e_{b\alpha}\right),
\end{equation}
where $\Gamma^{\alpha}_{\mu\nu}$ is the Christoffel symbol. For
the above metric, the only non-vanishing
$\Gamma^{\alpha}_{\mu\nu}$ are,
\begin{eqnarray}
\Gamma^{\theta}_{\phi\phi}=-b^{2}\sin\theta\ \cos\theta\ \ ;\ \
\Gamma^{\phi}_{\theta\phi}=\cot\theta.
\end{eqnarray}
Choosing the zweibein to be
\begin{eqnarray}
{e_{\mu}}^{a}={\rm diag}(1,b\sin\theta)\ \ ;\ \ {e_{a}}^{\mu}={\rm
diag}(1,1/b\sin\theta),
\end{eqnarray}
and the Dirac matrices,
\begin{equation}
\gamma^{\theta}=\sigma^{1}\ \ \ ;\ \ \ \gamma^{\phi}=\sigma^{2},
\end{equation}
the spin connection are found to be
\begin{eqnarray}
\Gamma_{\theta}=0\ \ ;\ \ \Gamma_{\phi}=-\frac{i}{2}b\cos\theta\
\sigma^{3}.
\end{eqnarray}
Here $\sigma^{i}$ are the Pauli matrices.

Now the Dirac operator can be written down explicitly as,
\begin{equation}
\left[\sigma^{1}\partial_{\theta}+\sigma^{2}\frac{1}{b\sin\theta}
\left(\partial_{\phi}+\Gamma_{\phi}\right)\right]\psi
=\left[\sigma^{1}\left(\partial_{\theta}+\frac{1}{2}\cot\theta\right)
+\sigma^{2}\frac{1}{b\sin\theta}\partial_{\phi}\right]\psi.
\label{2eigen}
\end{equation}
Suppose we write the eigenvalue of this operator as $\pm
i\kappa$ and  express the fermion field $\psi$ in two component form:
\beq
\psi=\left(\begin{array}{c}
\psi_{+}\\
\psi_{-}
\end{array}
\right)\,.
\eeq
Then we find  the following set of equations:
\begin{equation}
\left[\sigma^{1}\left(\partial_{\theta}+\frac{1}{2}\cot\theta\right)
+\sigma^{2}\frac{1}{b\sin\theta}\partial_{\phi}\right]\psi_{\pm}=
\pm i\kappa\psi_{\pm}.
\end{equation}
Let us consider $\psi_{+}$, where $\psi_{-}$ can be dealt with
analogously. Consider the equation for $\partial_{\phi}$:
\begin{equation}
\partial_{\phi}\chi^{(\pm)}_{m}=\pm im\chi^{(\pm)}_{m},
\end{equation}
that is,
\begin{equation}
\chi^{(\pm)}=e^{\pm im\phi}.
\end{equation}
Note that for spinors, one should get a sign change for a 2$\pi$
rotation in $\phi$. Therefore, the eigenvalues of $m$ should be
half-integers,
\begin{equation}
m=\frac{1}{2},\frac{3}{2},\frac{5}{2},\cdots.
\end{equation}
At this point it may be worth mentioning that we can also obtain the scalar angular eigenvalue for the 2-sphere by assuming that $m$ takes only integer values, i.e., $m=0,1,2,\dots$.

Returning to the eigenvalue equation  for $\psi_+$ we can make the following spinor separation of variables ansatz:
\begin{equation}
\psi^{(\pm)}_{+nm}=\left(
\begin{array}{c}
A^{(\pm)}_{n}(\theta)\chi^{(\pm)}_{m}(\phi) \\
B_{n}^{(\pm)}(\theta)\chi^{(\pm)}_{m}(\phi)
\end{array}
\right)\,,
\label{PMpsi}
\end{equation}
Putting this into the eigenvalue equation, we have the following
set of equations,
\begin{eqnarray}
&&\left(\partial_{\theta}+\frac{1}{2}\cot\theta\right)A_n^{(\pm)}
\mp\frac{m}{b\sin\theta}A_n^{(\pm)}=i\kappa B_n^{(\pm)},\nonumber\\
&&\left(\partial_{\theta}+\frac{1}{2}\cot\theta\right)B^{(\pm)}
\pm\frac{m}{b\sin\theta}B_n^{(\pm)}=i\kappa A_n^{(\pm)}.
\end{eqnarray}
These can be turned into second order equations. For $A_n^{(+)}$, we
have,
\begin{equation}
\left[\left(\partial_{\theta}+\frac{1}{2}\cot\theta\right)
+\frac{m}{b\sin\theta}\right]\left[\left(\partial_{\theta}+\frac{1}{2}\cot\theta\right)
-\frac{m}{b\sin\theta}\right]A_n^{(+)}=-\kappa^{2}A_n^{(+)}.
\end{equation}
One can get rid of the first derivative term by defining
\begin{equation}
A^{(+)}=\left(\sin\theta\right)^{-1/2}u.
\end{equation}
Then the equation for $u$ is
\begin{equation}
\frac{d^{2}u}{d\theta^{2}}+\left[\frac{1-4\left(\frac{m}{b}+\frac{1}{2}\right)^{2}}
{16\sin^{2}\frac{\theta}{2}}+\frac{1-4\left(\frac{m}{b}-\frac{1}{2}\right)^{2}}
{16\cos^{2}\frac{\theta}{2}}+\lambda^{2} \right]u=0.
\end{equation}
This equation is just the one for the Jacobi polynomial with the
solution
\begin{equation}
u=\left(\sin\frac{\theta}{2}\right)^{\alpha+\frac{1}{2}}
\left(\cos\frac{\theta}{2}\right)^{\beta+\frac{1}{2}}P_{n}^{(\alpha,\beta)}(\cos\theta),
\end{equation}
where $P_{n}^{(\alpha,\beta)}(\cos\theta)$ is the Jacobi
polynomial with
\begin{equation}
\alpha=\frac{m}{b}+\frac{1}{2}\ \ \ ;\ \ \
\beta=\frac{m}{b}-\frac{1}{2},
\end{equation}
and the eigenvalue obtained is
\begin{equation}
\kappa(n,m)=n+\frac{m}{b}+\frac{1}{2}
\end{equation}
where $n=0,1,2,\dots$ and $m=1/2,3/2,5/2,\dots$. To ensure
convergence $n$ must be an integer and thus
$P_{n}^{(\alpha,\beta)}$ is a polynomial.

For the regular two sphere, with $b=1$, we see that
\begin{equation}
\lambda=n+m+\frac{1}{2}=n'+1,
\end{equation}
where in the second step we have defined $n'=0,1,2,\dots$ with the constraint $m=\pm 1/2,\pm 3/2, \dots, \pm (n'+1/2)$, which is the expected result \cite{Camporesi}.
Thus, we can express the deficit eigenvalue as
\beq
\kappa(n',m)=n+|m|+\frac 1 2+|m|\left({1\over b}-1\right)=n'+|m|\left({1\over b}-1\right)
\eeq
where we have generalized to include positive and negative $m$.

It is now straightforward to generalize to $(d-2)$-dimensions, because we know the result when $b=1$ and hence can use eigenflow arguments similar to \cite{Siopsis}.\footnote{We have also confirmed this by checking the eigenvalue for the deficit 3-sphere.} Hence we find after dropping the prime:
\beq
\kappa(d,m)=n+\frac{d-2}{2}+|m|\left({1\over b}-1\right)\,.
\eeq
For $b=1$ the standard result for the regular $(d-2)$-sphere is obtained \cite{Camporesi,CCDN}.

\section{QNMs Using The Iyer and Will Method}

\begin{table}
\caption{Massless bulk Dirac fundamental ($p=0$) QNM frequencies (${\rm Re}(E)>0$) for a $d=6$ tensional BH plotted for various $(b,m,n)$ with $\mu=2$.}
\label{Dirac}
\begin{ruledtabular}
\begin{tabular}{l|llll}
$b$&    $n=0$&  $n=1$&  $n=2$&  $n=3$\\ ~&~&$~~~~m=1/2$&~&~\\
\hline 1     &0.79441-0.39649 i   &1.30313-0.38936 i
&1.77861-0.39134 i   &2.24148-0.39212 i\\ 0.9 &0.79563-0.38139 i
&1.28425-0.37603 i  &1.74227-0.37791 i  &2.18878-0.3786 i\\ 0.7
&0.80602-0.34805 i   &1.24916-0.34612 i   &1.66789-0.3477 i
&2.07762-0.34822 i\\ 0.5 &0.83723-0.30914 i  &1.22496-0.30991 i
&1.59608-0.31101 i  &1.96117-0.31133 i\\ 0.3 &0.92644-0.26089 i
&1.24267-0.26211 i  &1.55178-0.26253 i  &1.8582-0.26263 i\\ 0.1
&1.35902-0.18211 i  &1.57048-0.18211 i  &1.7816-0.18211 i
&1.9925-0.1821 i\\ 0.01    &5.02563-0.08452 i  &5.12323-0.08452 i
&5.22083-0.08452 i  &5.31843-0.08452 i\\ 0.001   &22.71727-0.03923
i &22.76257-0.03923 i &22.80787-0.03923 i &22.85317-0.03923
i\\\hline ~&~&~~~~$m=3/2$&~&~\\\hline 1      &  &1.30313-0.38936 i
&1.77861-0.39134 i  &2.24148-0.39212 i\\ 0.9 &  &1.33616-0.37626 i
&1.79225-0.37803 i  &2.23806-0.37864 i\\ 0.7 & &1.43038-0.34695 i
&1.84413-0.34801 i  &2.25202-0.3483 i\\ 0.5 & &1.59608-0.31101 i
&1.96117-0.31133 i &2.32389-0.3114 i\\ 0.3 & &1.96004-0.26264 i
&2.26503-0.26265 i &2.56951-0.26264 i\\ 0.1 & &3.46647-0.18209 i
&3.67689-0.18209 i &3.88729-0.18209 i\\ 0.01 & &14.78521-0.08452 i
&14.8828-0.08452 i &14.9804-0.08452 i\\ 0.001 & &68.01597-0.03923
i &68.06127-0.03923 i &68.10657-0.03923 i\\
\end{tabular}
\end{ruledtabular}
\end{table}

\par
\begin{figure}
\centering
\includegraphics[scale=.6]{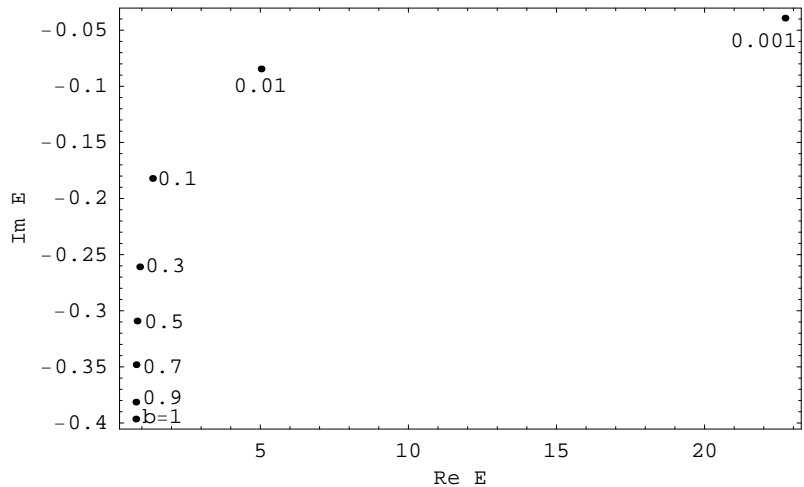}
\caption{Plot of the fundamental QNM ($p=0$, $n=0$, $m=1/2$) for varying tension, $b$. We see that as the tension increases the imaginary part vanishes.}
\label{fig:fundQNM}
\end{figure}

 To evaluate the QNM frequencies we adopt the WKB approximation developed by Iyer and Will \cite{IW}, also see references therein. Note that this analytic method has been used extensively in various BH cases \cite{Iyer}, where comparisons with other numerical results have been found to be accurate up to around 1\% for both the real and the imaginary parts of the frequencies for low-lying modes with $p<n$ (where $p$ is the mode number and $n$ is the spinor angular momentum quantum number). Furthermore, we have also included the $p=n=0$ modes in our results, shown in Table \ref{Dirac}. In figure \ref{fig:fundQNM}, we have plotted the fundamental QNM as a function of the tension, $b$.\\

The formula for the complex quasi-normal mode frequencies $E$ in the WKB approximation, carried to third order beyond the eikonal approximation, is given by \cite{IW}:
\begin{equation}
E^{2}=[V_{0}+(-2V^{''}_{0})^{1/2}\Lambda]
-i(p+\frac{1}{2})(-2V^{''}_{0})^{1/2}(1+\Omega) ,
\label{WKBeq}
\end{equation}
where we denote $V_0$ as the maximum of $V_1$ and
\begin{eqnarray}
\Lambda&=&\frac{1}{(-2V^{''}_{0})^{1/2}}
\left\{\frac{1}{8}\left(\frac{V_{0}^{(4)}}{V_{0}^{''}}\right)
\left(\frac{1}{4}+\alpha^{2}\right)- \frac{1}{288}
\left(\frac{V_{0}^{'''}}{V_{0}^{''}}\right)^{2}(7+60\alpha^{2})\right\},\\
\Omega&=&\frac{1}{(-2V_{0}^{''})}\left\{\frac{5}{6912}
\left(\frac{V_{0}^{'''}}{V_{0}^{''}}\right)^{4}(77+188\alpha^{2})
-\frac{1}{384}\left(\frac{{V_{0}^{'''}}^{2}V_{0}^{(4)}}{{V_{0}^{''}}^{3}}\right)
(51+100\alpha^{2})\right.\nonumber\\ &&\ \ \
+\frac{1}{2304}\left(\frac{V_{0}^{(4)}}{V_{0}^{''}}\right)^{2}(67+68\alpha^{2})
+\frac{1}{288}\left(\frac{V_{0}^{'''}V_{0}^{(5)}}{{V_{0}^{''}}^{2}}\right)(19+28\alpha^{2})
\left.
-\frac{1}{288}\left(\frac{V_{0}^{(6)}}{V_{0}^{''}}\right)(5+4\alpha^{2})\right\} .
\end{eqnarray}
Here
\begin{eqnarray}\label{eqn:alpha}
\alpha&=&p+\frac{1}{2},\ p=\left\{
\begin{array}{l}
0,1,2,\cdots,\ {\rm Re}(E)>0\\ -1,-2,-3,\cdots,\ {\rm Re}(E)<0
\end{array}
\right.
\qquad \mathrm{and} \qquad V_{0}^{(n)}=\left.\frac{d^{n}V}{dr_{\ast}^{n}}\right|_{r_{\ast}=r_{\ast}(r_{max})} .
\end{eqnarray}
\section{Large Angular Momentum \& Large Tension Limit}
\label{sec:tension}

\par A useful check of our numerical results is the exact analytic expression that can be obtained in the large overtone limit (also see \cite{Siopsis} for a similar discussion). If we now focus on the large angular momentum limit ($\kappa\to\infty$) we can easily extract an analytic expression for the QNMs to first order:
\beq E^{2}\approx V_{0} -i(p+\frac{1}{2})(-2V^{''}_{0})^{1/2} +
\dots , \label{largeK} \eeq where $V_0$ is the maximum of the
potential $V_1$, see equation (\ref{V1}). In the large $\kappa$
limit this potential now takes the form: \beq V_{1}
\Big|_{\kappa\to\infty} \approx {\kappa^2
r^{d-3}(r^{d-3}-r^{d-3}_H) \over r^{2(d-2)}} \,, \eeq where the
maximum of the potential in such a limit is then found to be: \beq
V_0 \Big|_{\kappa\to\infty} \approx \left(d-1\over
2\right)^{1\over d-3}\, r_H . \eeq In this case we find from the
1st order WKB approximation that: \beq E\Big|_{\kappa\to\infty}
\approx {2^{1\over{d-3}}\sqrt{d-3}\over (d-1)^{d-1\over
2(d-3)}r_H} \left[\kappa  - i \big(p+\frac 1 2\big)\sqrt{d-3}
\right] \,, \eeq where this result agrees with the our previous
result when $b=1$ \cite{CCDN}.\footnote{Note this is similar to
the scalar field result \cite{Kono}, but not identical.}

Interestingly, choosing $d=6$, and taking the limit $b\to 0$ leads
to an expression independent of $n$ (assuming $m\neq 0$ which is
the case for spin half fields in six dimensions) \beq
E\Big|_{b\to0,\, d=6} \approx {\sqrt{3}\,b^{1/3}\over (5)^{5\over
6}} \left[{m\over b}  - i \big(p+\frac 1 2\big)\sqrt{3} \right]
\,, \eeq where we have used the fact that $r_H= (\mu/b)^{1/3}$
with $\mu=2$. Thus, as $b\to 0$ we see that the imaginary part
becomes negligible (for fixed overtone $p$) and this agrees with
our plot in Figure 1. Similar analysis in \cite{Siopsis} shows
that this limit is independent of the type of perturbation
exciting the BH (to lowest order in an inverse power series in
$\kappa$).

\section{Concluding remarks}

\par We have investigated the effect of brane tension on the low lying massless Dirac QNMs of a BH localized on a tense 3-brane in six dimensions. Conformal techniques were used to separate the time-radial and angular parts of the Dirac field analogously to the methods used in the zero tension case \cite{CCDN}. We have found that the only difference in the calculation between the tense and tensionless case appears in the value of the angular eigenvalues, $\kappa$, which now depends on the the amount of tension, $b$. By calculating the unphysical deficit 2-sphere eigenvalue, $\kappa$, we were able to determine the physically relevant 4-sphere eigenvalue by induction. We then computed the low lying QNMs for various tension, $b$, using the 3rd order WKB approximation \cite{Iyer}, see Table \ref{Dirac}. As a byproduct of our method, this also gives a rigorous proof of the result found in \cite{Siopsis} for integer spin fields on the deficit 2-sphere with no assumptions made on the form of $b$, when we choose integer values of $m$.

\par As can be seen from the plot of the fundamental mode in figure \ref{fig:fundQNM} a general feature of these QNMs is that the imaginary part disappears with increasing tension ($b\rightarrow 0$). Furthermore, as the tension is increased the real part grows without bound. These results agree with the large tension limit that was calculated analytically in section \ref{sec:tension}. Because the imaginary part gets smaller for larger tensions the amount of energy available for other bulk processes diminishes, which for example has been demonstrated in \cite{Dai} for bulk BH emission rates for integer spin fields. Note, brane-localized QNMs/emissions are not affected by the brane tension, for a fixed horizon.

\par Interestingly, the QNMs in the large $p$ asymptotic limit can be calculated with tension using the method by Andersson and Howls \cite{Andersson}, who have combined the WKB formalism with the monodromy method of Motl and Neitzke \cite{Motl}, also see \cite{Cho2}. The calculation in the tense brane case follows exactly that of the tenseless case \cite{CCDN}. Thus in the large $p$ limit the QNM is purely imaginary with value, $E_p=-i2\pi T_H p$. However, since the temperature is related to the Schwarzschild radius through the equation $T_H=(d-3)/4\pi r_H$ and the radius is altered by the tension via equation (\ref{eqn:tenseradius}), any tension will reduce the absolute value of the QNM in the large $p$ asymptotic limit.

\par In conclusion it is worth emphasising that tension is a fundamental property of branes in higher dimensional brane world models, yet until recent times, understanding the phenomenological consequences of tension on BHs has not been possible. Such knowledge is essential in order to determine realistic BH production rates and observational signatures at the LHC. In a further work we will investigate the Hawking emission for bulk Dirac fields from a BH embedding in a codimension two tense brane in six dimensions to compare these results with other fields on such a background \cite{Dai} and with fermions in the tensionless case \cite{CCDN2}.

\acknowledgments

\par HTC was supported in part by the National Science Council of the Republic of China under the Grant NSC 96-2112-M-032-006-MY3, and the National Center for Theoretical Sciences. JD wishes to thank Dr. G. C. Joshi for his advice and supervision during the production of this work.


\end{document}